\documentclass[11pt]{article}

\usepackage{pb-diagram}
\usepackage{latexsym}
\usepackage{amsfonts}
\usepackage[usenames,dvipsnames]{xcolor}
\usepackage{graphicx,amssymb,amsmath,epsfig}
\usepackage[english]{babel}
\usepackage{graphicx}
\usepackage{dcolumn}
\usepackage{bm}
\usepackage{slashed}

\definecolor{myblue}{rgb}{0,0,0.8}
\usepackage[colorlinks=true
,urlcolor=myblue	
,anchorcolor=myblue
,citecolor=myblue
,filecolor=myblue
,linkcolor=myblue
,menucolor=myblue
,pagecolor=myblue
,linktocpage=true  
]{hyperref}

\catcode`\@=11
\def\marginnote#1{}

\newcount\hour
\newcount\minute
\newtoks\amorpm
\hour=\time\divide\hour by60 \minute=\time{\multiply\hour by60
\global\advance\minute by-\hour}\edef\standardtime{{\ifnum\hour<12
\global\amorpm={am}%
        \else\global\amorpm={pm}\advance\hour by-12 \fi
        \ifnum\hour=0 \hour=12 \fi
        \number\hour:\ifnum\minute<10
0\fi\number\minute\the\amorpm}}
\edef\militarytime{\number\hour:\ifnum\minute<10 0\fi\number\minute}

\def\draftlabel#1{{\@bsphack\if@filesw {\let\thepage\relax
   \xdef\@gtempa{\write\@auxout{\string
      \newlabel{#1}{{\@currentlabel}{\thepage}}}}}\@gtempa
   \if@nobreak \ifvmode\nobreak\fi\fi\fi\@esphack}
        \gdef\@eqnlabel{#1}}
\def\@eqnlabel{}
\def\@vacuum{}
\def\draftmarginnote#1{\marginpar{\raggedright\scriptsize\tt#1}}
\def\draft{\oddsidemargin -.5truein
        \def\@oddfoot{\sl preliminary draft \hfil
        \rm\thepage\hfil\sl\today\quad\militarytime}
        \let\@evenfoot\@oddfoot \overfullrule 3pt
        \let\label=\draftlabel
        \let\marginnote=\draftmarginnote

\def\@eqnnum{(\theequation)\rlap{\kern\marginparsep\tt\@eqnlabel}%
\global\let\@eqnlabel\@vacuum}  }

\def\numberbysection{\@addtoreset{equation}{section}
        \def\theequation{\thesection.\arabic{equation}}}

\def\underline#1{\relax\ifmmode\@@underline#1\else
 $\@@underline{\hbox{#1}}$\relax\fi}

\catcode`@=12 \relax

\numberbysection

\topmargin by -\headsep
\newcommand{\bi}{\begin{itemize}}
\newcommand{\ei}{\end{itemize}}
\newcommand{\bc}{\begin{center}}
\newcommand{\ec}{\end{center}}

\newcommand{\be}{\begin{equation}}
\newcommand{\ee}{\end{equation}}

\newcommand{\bqn}{\begin{eqnarray}}
\newcommand{\eqn}{\end{eqnarray}}

\def\br{\begin{eqnarray}}
\def\er{\end{eqnarray}}

\def\({\left(}
\def\){\right)}
\def\[{\left[}
\def\]{\right]}

\def\sl{\sqrt{\lambda}}

\def\ba{\begin{align}}
\def\ea{\end{align}}
\def\be{\begin{eqnarray}}
\def\ee{\end{eqnarray}}

\begin{document}

\vspace*{1cm}
\noindent

\vskip 1 cm
\begin{center}
{\Large\bf A note on one-loop soliton quantum mass corrections}
\end{center}
\normalsize
\vskip 1cm
\begin{center}
{A. R. Aguirre}\footnote{\href{mailto:alexis.roaaguirre@unifei.edu.br}{alexis.roaaguirre@unifei.edu.br}} and  G. Flores-Hidalgo\footnote{\href{mailto:gfloreshidalgo@unifei.edu.br}{gfloreshidalgo@unifei.edu.br}}\\[.7cm]

\par \vskip .1in \noindent
\emph{Institute of Physics and Chemistry, Federal University of Itajub\'a (UNIFEI),\\
Itajub\'a, Minas Gerais, 37500-903, Brazil}
\vskip 2cm

\end{center}

\begin{abstract}

We develop an alternative derivation of the renormalized expression for the
one-loop soliton quantum mass corrections in $(1 + 1)$-dimensional scalar field theories. We regularize implicitly
such quantity by subtracting and adding its corresponding tadpole graph contribution and use the
renormalization prescription that such added term vanishes with adequate counterterms. As a result,
we get a finite unambiguous formula for the soliton quantum mass corrections up to one-loop order,
which turns to be independent of the chosen regularization scheme.

\end{abstract}

\newpage



\section{Introduction}
\label{sec:intro}

There has been a great deal of progress in studying one loop quantum corrections to the kink mass in (1+1)-dimensional field 
theories since the first appearance of these calculations in the bosonic $\phi^4$ and sine-Gordon models  \cite{Dashen}--\cite{Faddeev}. 
After that, several authors considered the corresponding supersymmetric extensions of those models and since then  different 
approaches have been developed  to calculate quantum corrections to the supersymmetric kink mass and central charge \cite{Dadda}--\cite{Imbimbo}. For some time, the two main concerns were if the bosonic and fermionic contributions in the quantum 
corrections to the supersymmetric kink mass cancel each other, and if the BPS saturation condition survives at the
quantum level, which were exhaustively investigated in a  series of publications \cite{Elizalde}--\cite{Rebhan09}.

Recently, these issues have been reconsidered by introducing new tools and methods to deal with renormalization, regularization
 and calculation of the one loop corrections to the kink mass in bosonic scalar and supersymmetric field theories
\cite{gousheh2012}--\cite{Gabriel16}, which bring up new interesting insights on this topic.

The purpose of this short note is to provide a simple derivation of the renormalized formula for the one loop soliton quantum mass correction in (1+1) dimensional bosonic field theory. As the result, we will get a formula, Eq. (\ref{meven}), which turns out to be independent of the regularization scheme used.


\section{The method}

Let us start with the bosonic Lagrangian,
\bqn
{\cal L}&=&\frac{1}{2}\partial_\mu\phi\partial^\mu \phi-U(\phi)+\delta {\cal L},
\label{redefined}
\eqn
where the potential $U(\phi)$ has at least two degenerate minima, and $\delta {\cal L}$ contains adequate  counterterms in order to
render finite the theory.  By quantizing around the static kink configuration $\phi_c$, we can write the soliton mass at one loop order in the following form,
\bqn
M&=& E[\phi_c]+\delta {M}+\frac{1}{2}\sum_{n}\omega_{n}
-\frac{1}{2}\sum_{k}\omega(k)
\;,
\label{e1}
\eqn
where $E[\phi_c]$ is the energy of the static classical kink configuration, $\delta M$ are the counterterm  contributions
from $\delta {\cal L}$ term at one loop order and
the eigenfrequencies  $\omega_{n}$ and $\omega(k)$ are given by the eigenvalue equations 
\bqn
\left[-\frac{d^2}{dx^2}+U''[\phi_c(x)]\right]\eta_n&=&\omega_{n}^2\eta_n,
\label{bosef1}
\eqn
and
\bqn
\left[-\frac{d^2}{dx^2}+m^2 \right]\eta_k &=&\omega^2(k)\eta_k,
\label{freeboson}
\eqn
with $m$ being the mass of the quantum fluctuations around the trivial vacua. The last two terms in (\ref{e1}) usually are  logarithmically divergent, which requires the use of certain regularization and renormalization schemes. 
There are several  regularization techniques that have been used to deal with this problem (see for instance Refs. \cite{Dashen}, \cite{Chan}--\cite{Weigel17}, and references therein). 

In this note we consider a simple method to regularize the last two terms in Eq. (\ref{e1}). The method is based on the following formal identity \cite{boya},
\bqn
\frac{1}{2}\left[\sum_{n}\omega_n-\sum_k\omega(k)\right] &=&
\frac{1}{2}\int_{-\infty}^\infty\frac{d\omega}{2\pi}\,{\rm Tr}\ln
\left(1+\hat{A}\right),
\label{eaction1}
\eqn
where 
\bqn
\hat{A} &=& \frac{{\cal V}}{\omega^2-\frac{d^2}{dx^2}+m^2},
\label{operator}
\eqn
with ${\cal V}=U''[\phi_c(x)]-m^2$. Equation (\ref{eaction1}) can be expanded formally in
powers of $\hat{A}$, and then, in terms of Feynman graphs, we have
\bqn
\frac{1}{2}\left[\sum_{n}\omega_n-\sum_k\omega(k)\right] &=&\hspace{0.2cm}\parbox[b]{7.5cm}{ \raisebox{-0.95cm}
{\psfig{file=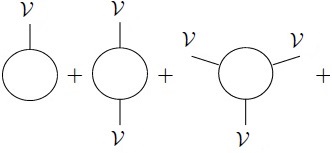,height=2.1cm,width=4.5cm}}} \hspace{-2.8cm}\dots \,,
\label{eaction2}
\eqn
from which we can identify the tadpole graph  contribution to the one loop soliton mass,
as the only ultraviolet divergent graph, and whose expression is 
given by
\bqn
\parbox[b]{2cm}{ \raisebox{-0.3cm}{\psfig{file=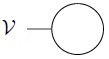,
height=0.85cm,width=1.55cm}}}\quad\hspace{-0.7cm}
&=& \frac{1}{2}\!\int_{-\infty}^\infty\! \frac{d\omega}{2\pi} \,{\rm Tr}\hat{A}\;.
\label{tadpole}
\eqn 
Therefore, by adding and subtracting  the tadpole graph above in (\ref{e1}), and using the renormalization prescription
that the added tadpole graph cancels with $\delta M$, we get the renormalized
finite result for the one loop soliton mass,
\bqn
M &=& E[\phi_c]+\frac{1}{2}\sum_n\!\omega_n-\frac{1}{2}\sum_k\! \omega(k)
- \frac{1}{2}\int_{-\infty}^\infty\! \frac{d\omega}{2\pi} {\rm Tr}\hat{A}\;.
\label{renormalized}
\eqn
Of course,  all the above manipulations are meaningful only if  the divergent terms are taken as already regularized. 
Consequently, in order to get finite unambiguous results through (\ref{renormalized}), we have to consider the last
three terms as a whole, i.e. we have to use same
regularization procedure when dealing with each divergent term. 
 Thus, considering such quantities as implicitly regularized,
we have to take due care only
in passing 
from formal sums to integrals for the continuous spectrum given by Eqs. (\ref{bosef1}) and (\ref{freeboson}). 
For this end we note that
 the free-soliton eigenfrequencies in Eq. (\ref{freeboson}) are continuous and given by $\omega(k)=\sqrt{k^2+m^2}$. On the other hand, the soliton eigenfrequencies in Eq. (\ref{bosef1}) fall in two sets, a finite discrete set denoted by $\omega_{i}$, and a continuous set  $\omega(q)=\sqrt{q^2+m^2}$. Both  $\omega(k)$ and
$\omega(q)$, range  from $m$ to $\infty$, but they do not cancel each other in Eq. (\ref{e1})  because their corresponding densities of states are different. Splitting the second sum in (\ref{e1}), we have
\bqn
M &=&E[\phi_c]+\frac{1}{2}\sum_{i=1}^N\omega_i+\Delta\;,
\label{split}
\eqn
with $\Delta$ given by
\begin{eqnarray}
\Delta &=&\frac{1}{2} \sum_{q=-\infty}^\infty \omega(q)
-\frac{1}{2} \sum_{k=-\infty}^\infty \omega(k)- \frac{1}{2}\int_{-\infty}^\infty \frac{d\omega}{2\pi} {\rm Tr}\hat{A}.
\label{a2}
\end{eqnarray}
In order to go from formal sums to integrals
we enclose the system in a box of size $L$, and then taking the limit $L\to\infty$ after applying periodic boundary conditions. For the free 
eigenfunctions $\eta_k\propto { e}^{ikx}$, we have the condition ${e}^{-ikL/2}={ e}
^{ikL/2}$, and then
\bqn
 k_n &=& \frac{2\pi n}{L}, \qquad n=0,\pm 1,\pm 2,...
 \label{kfree}
\eqn
In this case we found that the free density of states is given by,
\br 
\frac{1}{(k_{n+1}-k_n)} &=&\frac{L}{(2\pi)}.
\er 
For the
soliton eigenfrequencies $\omega(q)$,  let us consider first the simplest case in which ${\cal V}(x)$ is a reflectionless 
potential, which is the case of the sine-Gordon and $\phi^4$ models. In this case, the asymptotic behaviour of $\eta_q(x)$ is 
given by,
\bqn
\eta_q(x)  &=&\left\{\begin{array}{ll}
{ e}^{iqx} &x\to-\infty\\
{ e}^{iqx+i\delta(q)} &x\to+\infty
\end{array}\right.\;,
\label{assymp}
\eqn
where $\delta(q)$ is the scattering  phase shift. Imposing periodic boundary conditions, $\eta_q(-L/2)=\eta_q(L/2)$, we get
\br
q_n &=&\frac{2\pi n}{L}-\frac{\delta(q_n)}{L},~~n=0,\pm1,\pm2,\dots \;.
\label{ad3}
\er
Now, by using the free-soliton basis $\{ \langle x|k_n\rangle=e^{ik_nx}/\sqrt{L}\}$ to compute the trace  in the last term of Eq. (\ref{a2}) we get after integration in $\omega$,
\begin{eqnarray}
\int_{-\infty}^\infty \frac{d\omega}{4\pi} {\rm Tr}\hat{A}&=&
\int_{-\infty}^\infty\frac{d\omega}{4\pi} \sum_{n=-\infty}^\infty\langle k_n|\frac{{ \cal V}(x)}{\omega^2-\frac{d}{dx^2}+m^2}
|k_n \rangle,\nonumber\\
&=&\frac{\langle {\cal V}\rangle}{4}\sum_{n=-\infty}^\infty \frac{1}{L\sqrt{k_n^2+m^2}}
\;,
\label{tadpoleb}
\end{eqnarray}
with
\bqn
\langle {\cal V}\rangle &=&\int_{-\infty}^\infty {\cal V}(x) dx\;.
\label{media}
\eqn
Therefore, { by using that $\omega(-q)=\omega(q)$ (similarly for $\omega(k)$),  Eq. (\ref{a2}) can be
rewritten as follows,}
\br
{\Delta} &=&{\sum_{q=0}^\infty \omega(q)
-\sum_{k=0}^\infty \omega(k)- \frac{1}{2}\int_{-\infty}^\infty \frac{d\omega}{2\pi} {\rm Tr}\hat{A}, }   \nonumber \\
&=&\sum_{n=n_0}^{\infty}\sqrt{q_n^2+m^2}
-\sum_{n=0}^{\infty}\sqrt{k_n^2+m^2}
-\frac{\langle {\cal V}\rangle}{4}\sum_{n=-\infty}^\infty \frac{1}{L\sqrt{k_n^2+m^2}}
\;,
\label{a6}
\er
where $n_0$ in the first sum is not necessarily equal
to zero, but must be chosen as the correct integer corresponding to $q=0$ according to Eq. (\ref{ad3}). 
So, by setting \mbox{$q_{n_0}=0^+$} in
Eq. (\ref{ad3}) and using the one dimensional Levinson theorem {for potential with half-bound states (which is the case of the reflectionless potentials and the free case)\cite{barton}}, \mbox{$\delta(0^+)=N\pi$},  we have
\br 
2\pi n_0/L- N \pi/L&=&0,
\er
from which we find that $n_0= N/2$.  It is worth noting that this solution makes sense only
if $ N$ is even, which will be assumed at first. Now, by setting 
$N=2 l$, we have  that $n_0=l$, and Eq. (\ref{a6}) can be written as 
\bqn
 \Delta&=&
\sum_{n=l}^\infty\sqrt{q_n^2+m^2}-\sum_{n=0}^\infty\sqrt{k_n^2+m^2}
-\frac{\langle {\cal V}\rangle}{4}\sum_{n=-\infty}^\infty \frac{1}{L\sqrt{k_n^2+m^2}}\nonumber\\
&+&\sum_{n=0}^\infty\left(\sqrt{q_n^2+m^2}-\sqrt{k_n^2+m^2}\right)
-\sum_{n=0}^{l-1}\sqrt{q_n^2+m^2}-\frac{\langle {\cal V}\rangle}{4}\!\sum_{n=-\infty}^\infty\! \frac{1}{L\sqrt{k_n^2+m^2}}\;.
\label{s1}
\eqn
From Eqs.  (\ref{kfree}) and (\ref{ad3}), we obtain
\bqn
\sqrt{q_n^2+m^2}&=&\sqrt{k_n^2+m^2}-\frac{k_n \,\delta(k_n)}{L
\sqrt{k_n^2+m^2}}+{\cal O}\!\left(L^{-2}\right)\!.
\label{s2}
\eqn
Substituting in (\ref{s1}), we have
\begin{eqnarray}
\Delta&=&-\sum_{n=0}^\infty
\frac{k_n\delta(k_n)}{L\sqrt{k_n^2+m^2}}-\sum_{n=0}^{l-1}\sqrt{q_n^2+m^2}
-\frac{\langle {\cal V}\rangle}{4}\sum_{n=-\infty}^\infty \frac{1}{L\sqrt{k_n^2+m^2}}+{\cal O}\!\left(L^{-2}\right).\quad\mbox{}
\label{s33}
\end{eqnarray}
Now, by replacing Eqs. (\ref{s33})  in Eq. (\ref{split}), and taking the  limit $L\to\infty$, we get 
\bqn
M&=&E[\phi_c]+\frac{1}{2}\sum_{i=1}^N(\omega_i  -m )
- \int_0^\infty \frac{dk}{4\pi}\frac{[2k\delta(k)+\langle  {\cal V}\rangle]}{\sqrt{k^2+m^2}},
\label{meven}
\eqn
where we have used that $q_0,q_1,..,q_{l-1}$, approach zero in the limit $L\to\infty$ in the {second sum} in (\ref{s33}). 
Equation (\ref{meven}) corresponds to  the renormalized formula for the one loop soliton quantum mass correction,
see for example \cite{Graham98,Livro}.

Let us now consider the case when $ N$ is odd. In this case $n_0=N/2$ will be a half integer, and then it will be necessary
to disregard this value since it does not satisfy the condition (\ref{ad3}). In fact, it turns to be that such  half integer 
$n_0$ satisfies anti-periodic boundary conditions. Now, disregarding $n_0$ is totally equivalent to disregard $q=0$,
and this implies now that the {\mbox{Eq. (\ref{a6})} } is no longer correct in the box. Then, by introducing a 
proper modification, we get  a new form of the expression {(\ref{a6}) } for the odd case, namely
\begin{eqnarray}
\Delta&=& {\frac{1}{2} \sum_{q=-\infty}^\infty \omega(q)
-\frac{1}{2} \sum_{k=-\infty}^\infty \omega(k)- \frac{1}{2}\int_{-\infty}^\infty \frac{d\omega}{2\pi} {\rm Tr}\hat{A} },\nonumber\\
&=&\sum_{q>0}\sqrt{q^2+m^2}-\sum_{k>0}\sqrt{k^2+m^2}-\frac{m}{2}
-\frac{\langle {\cal V}\rangle}{4}\sum_{n=-\infty}^\infty \frac{1}{L\sqrt{k_n^2+m^2}},\nonumber\\
&=&\sum_{n>n_0}^\infty\sqrt{q_n^2+m^2}-\sum_{n=1}^\infty\sqrt{k_n^2+m^2}
-\frac{m}{2}
-\frac{\langle {\cal V}\rangle}{4}\sum_{n=-\infty}^\infty \frac{1}{L\sqrt{k_n^2+m^2}},
\label{deltaodd}
\end{eqnarray}
{where we have used again the symmetry $\omega(q)=\omega(-q)$, $\omega(k)=\omega(-k)$, and that $\omega(0)=m$, by passing from the first to the second line}. The first sum in Eq (\ref{deltaodd}) must begin at the next integer greater than $n_0$. So, by setting $N=2l+1$,  we get
\bqn
\Delta &=&
\sum_{n=l+1}^\infty\sqrt{q_n^2+m^2}-\sum_{n=1}^\infty\sqrt{k_n^2+m^2}
-\frac{m}{2}
-\frac{\langle {\cal V}\rangle}{4}\sum_{n=-\infty}^\infty \frac{1}{L\sqrt{k_n^2+m^2}},\nonumber\\
&=&\sum_{n=1}^{\infty}\left(\sqrt{q_n^2+m^2}-\sqrt{k_n^2+m^2}\right)
-\sum_{n=1}^{l}\sqrt{q_n^2+m^2}
-\frac{m}{2}-\frac{\langle {\cal V}\rangle}{4}\sum_{n=-\infty}^\infty \frac{1}{L\sqrt{k_n^2+m^2}}.\quad\mbox{\quad}
\label{xx1}
\eqn
Now, by using Eq. (\ref{s2}), and  taking the  limit $L\to\infty$,  we find
\begin{eqnarray}
\Delta&=&-(2l+1)\frac{m}{2}-\int_{0}^{\infty}\frac{dk}{4\pi}\frac{[2k\delta(k)+\langle{\cal V}\rangle]}{\sqrt{k^2+m^2}},
\label{dodd}
\end{eqnarray}
and replacing this expression in (\ref{split}), we get the same formula  (\ref{meven}) of the even case.

{At this point it is worth highlighting that in its more general form, the Levinson's theorem in one dimension includes a one-half contribution, i.e $\delta(0) = (N-\frac{1}{2})\pi$. However, we can show that the results are not altered, and that the same procedure can be applied. Note that, in our method the Levinson's theorem is used to determine the lowest integer $n_0$, from where the sum over the soliton eigenfrequencies starts. In that case, we can see that the relation (19) is modified by,
\br 
 \frac{2\pi n_0}{L} - \frac{\pi}{L}\(N-\frac{1}{2}\)&=&0,  \qquad \longrightarrow\qquad n_0 = \frac{N}{2}-\frac{1}{4}. \nonumber
\er
Then, whether $N$ being even or odd, we should disregard that value (which is equivalent to disregard $q=0$ from the spectrum) since it does not satisfy the condition (15), and then the sum should start from the next integer. In the even case $N=2l$, we should start from $n_0 =l$, while in the odd case, $N=2l+1$, from $n_0=l+1$. After doing that, the derivation follows similar steps as the odd case.}

{On the other hand},  Eq. (\ref{meven})  can be also obtained for more general potentials ${\cal V}$, 
which are not necessarily reflectionless. In that case, it can be shown that identical results are obtained once the phase shift
 is properly generalized  in terms 
of the  $S$-matrix of the associated one-dimensional scattering problem given by the 
continuous solutions of Eq. (\ref{bosef1}). In terms of the reflection and 
transmission coefficient amplitudes, the $S$-matrix {can be parametrized in the following form},\cite{barton} 
{
\br
S &=& \left( \begin{array}{cc} T & R'\\R&T
\end{array}\right),
\er
where $R$ denotes the reflection coefficient of scattering from left to right, and $R'$ from right to the left \cite{Bianchi}. From unitarity of the $S$ matrix, we have $T^*R'+TR^*=0$, and then we obtain}
\begin{eqnarray}
S(k) &=&\left(
\begin{array}{lc}
T(k)&-R^\ast(k)T(k)/T^\ast(k)\\
R(k)&T(k)\end{array}
\right),\label{10}
\end{eqnarray}
{
Now, to decouple into the scattering channels, we can be performed an unitary transformation $USU^\dagger$, where
\br
U &=&\frac{1}{\sqrt{2}} \left[ \begin{array}{cc} e^{2i\gamma(k)} &-1\\1& e^{-2i\gamma(k)}
\end{array}\right], \qquad e^{2i\gamma(k)}=\sqrt{R'/R},
\er
and 
\br 
USU^\dagger &=& \left[ \begin{array}{cc} e^{2i\delta_1(k)} & 0\\0&e^{2i\delta_2(k)}
\end{array}\right].
\er
Then, the scattering decoupled channels become
\br
 \eta_1(x) &=& A_1 \sin\(kx+\gamma(k) \pm \delta_1(k)\), \qquad x\to \pm \infty, \nonumber\\
 \eta_2(x) &=& A_2 \cos\(kx+\gamma(k) \pm \delta_2(k)\), \qquad x\to \pm \infty.\nonumber
\er
By applying periodic boundary conditions at $x=\pm L/2$, we will find
\br
 k&=&\frac{2\pi n}{L} - \frac{2}{L}\delta_1(k), \qquad k\,\,=\,\,\frac{2\pi n}{L} - \frac{2}{L}\delta_2(k), \nonumber
\er
where each scattering phase shift channel fallows similar equation as eq. (15). Therefore,
repeating the same arguments above, we get the same formula (\ref{meven}) for the
one loop renormalized soliton mass, in terms of the total phase shift defined as
 $\delta(k) = \delta_1(k) +\delta_2(k)$. 
In addition, by taking the determinant of the $S$ matrix, we find that the phase shift is given by,}
\begin{eqnarray}
\delta(k)&=&\frac{1}{2i}\ln\det S(k),\nonumber\\
&=&\frac{1}{2i}\ln\left[\frac{T(k)}{T^\ast(k)}\right],\label{11}
\end{eqnarray}
{where we have used that $|T|^2+|R|^2=1$.} It is also worth mentioning that unitarity of the $S$-matrix  ensures that the phase shift $\delta(k)$ in  Eq. (\ref{11}) is a real function of $k$. 

\vspace{1cm}
\section{Discussion and conclusions}

The formula (\ref{meven}) that we have obtained for the renormalized mass of the soliton at one loop order is totally equivalent to the one obtained some time ago in reference \cite{Graham98}, where authors introduced the so called phase shift technique. At this point, we call attention to
the fact that although we have used the scattering phase shift as a basic ingredient, 
{our approach is  different
from Ref. \cite{Graham98}. First of all, in order to regularize the quantum mass
corrections, the authors subtracted the first  Born approximation from the phase shift,
and in order to overcome the infrared divergence introduced
by the first  Born approximation, the one dimensional Levinson's theorem is used. In our approach, we regularize implicitly the one loop soliton mass by substracting the
  tadpole graph contribution to it, which is free of infrared divergences. Therefore no use
  is made of the one dimensional Levinson's theorem with regard to the regularization issue.
 We get the same results, since we are using the same renormalization prescription.} 

From Eq. (\ref{s1}), we note that our method resembles somehow  the {so called} mode number regularization method.
However, we need to call attention again to the fact that the starting points and assumptions
 in both methods  are quite different.
 On one hand, in the mode number regularization method the system is enclosed in a
finite discrete box, which implies that the number of modes of Eqs. (\ref{bosef1}) and (\ref{freeboson}) become finite, and then they are assumed to be the same in both cases. On the other hand, in the method presented in this note
we enclose the system in a non-discrete box of finite size, and therefore the number of modes is no longer finite. So, in our approach, nothing else about the  number of modes is
then assumed. We can proceed in this way since, as it was already mentioned, $\Delta$ given by Eq. (\ref{s1}) is finite and  {unambiguous, if each separately divergent term is
considered as implicitly regularized in the same scheme}. Then, the only issue that we have to be careful about is in passing from formal sums to integrals for the continuum spectrum of Eqs. (\ref{bosef1}) and (\ref{freeboson}).

\newpage
\noindent
{\bf Acknowledgements} \\
\vskip .1cm \noindent
Authors would like to thank CAPES-Brazil for partial financial support. We are also very grateful to members of the Institute of Physics and Chemistry (IFQ-UNIFEI) for the propitious environment for developing this work. {The authors also thank the referees for useful comments and suggestions that helped us to improve the readability of our work, as well as to clarify some important issues.}

\vskip 1cm

\end{document}